\documentclass[aip,jap,numerical,twocolumn,groupedaddress, reprint, 10pt]{revtex4-1}
\usepackage{graphicx} 
\usepackage{color}

\usepackage[english]{babel}
\usepackage[utf8]{inputenc}

\usepackage{amssymb}
\usepackage{hyperref}

\draft % marks overfull lines with a black rule on the right

\begin{document}

% Use the \preprint command to place your local institutional report number 
% on the title page in preprint mode.
% Multiple \preprint commands are allowed.
%\preprint{}

\title{Graphene nanoribbons: relevance of etching process} %Title of paper

% repeat the \author .. \affiliation  etc. as needed
% \email, \thanks, \homepage, \altaffiliation all apply to the current author.
% Explanatory text should go in the []'s, 
% actual e-mail address or url should go in the {}'s for \email and \homepage.
% Please use the appropriate macro for the type of information

% \affiliation command applies to all authors since the last \affiliation command. 
% The \affiliation command should follow the other information.

\author{P. Simonet}
\email{psimonet@phys.ethz.ch}
\author{D. Bischoff}
\author{A. Moser}
\author{T. Ihn}
\author{K. Ensslin}
\affiliation{ Solid State Physics Laboratory, ETH Zurich, 8093 Zurich, Switzerland}
\date{\today}

\begin{abstract}
Most graphene nanoribbons in the experimental literature are patterned using plasma etching. Various etching processes induce different types of defects and do not necessarily result in the same electronic and structural ribbon properties. This study focuses on two frequently used etching techniques, namely O$_{\rm 2}$ plasma ashing and O$_{\rm 2} +$Ar reactive ion etching (RIE). O$_{\rm 2}$ plasma ashing represents an alternative to RIE physical etching for sensitive substrates, as it is a more gentle chemical process. We find that plasma ashing creates defective graphene in the exposed trenches, resulting in instabilities in the ribbon transport. These are probably caused by more or larger localized states at the edges of the ashed device compared to the RIE defined device.
\end{abstract}

%\pacs{71.15.Mb, 81.05.ue, 72.80.Vp}% insert suggested PACS numbers in braces on next line

\maketitle %\maketitle must follow title, authors, abstract and \pacs

% Body of paper goes here. Use proper sectioning commands. 
\section{Introduction}
Graphene's exceptional electronic properties have triggered an entire research field \cite{geim_graphene:_2009}. Its relativistic charge carriers experience little scattering making quantum phenomena observable at room temperature \cite{novoselov_room-temperature_2007}. Thus, a large effort of graphene research focuses on building quantum devices. Among these, graphene quantum dots are promising candidates for the implementation of spin-qubits, because long spin coherence times were predicted in these systems compared to usual semiconductors. The transport mechanisms in such nanostructures are, however, more complicated than expected \cite{han_energy_2007, ponomarenko_chaotic_2008}. Graphene nanoribbons are a fundamental component of any graphene nanodevice and have therefore been extensively studied. 
There are many ways to produce graphene nanoribbons, including  plasma etching \cite{han_energy_2007}, chemical processes to grow \cite{pan_topographic_2012, baringhaus_exceptional_2014} or define them \cite{li_chemically_2008,masubuchi_fabrication_2009,wang_etching_2010}, electrical techniques \cite{moser_fabrication_2009, allen_gate-defined_2012}, bottom-up fabrication \cite{cai_atomically_2010}, physical bombardment \cite{lemme_etching_2009, lu_situ_2011} and natural exfoliation \cite{adam_density_2008}. Graphene nanoribbons fabricated with many different techniques showed a suppressed conductance at low Fermi energies \cite{han_energy_2007, li_chemically_2008, masubuchi_fabrication_2009, wang_etching_2010, ki_crossover_2012, adam_density_2008}, which has been identified to originate from disorder-induced localized charges in the ribbons \cite{sols_coulomb_2007,todd_quantum_2008,molitor_transport_2009}. This suppressed transport was also observed in graphene devices where the bulk disorder was significantly reduced, indicating that edge disorder was sufficient to form such localized states in the ribbons \cite{bischoff_reactive-ion-etched_2012}. Understanding the origin of disorder in the ribbons and particularly at their edges constitutes a first step towards its elimination.

Plasma etching of lithographically patterned graphene is one of the most widely used techniques due to the possibility of patterning nearly arbitrary geometries. Moreover, it can be tuned from being a chemical to a physical process. Indeed, a variation of either the gas, pressure or plasma acceleration voltage changes the etching mechanism. If a low pressure gas of heavy atoms like argon is ionized in a chamber where the sample is placed on the biased electrode, such as in a reactive ion etching (RIE) chamber, then a physical and directed etching will occur \cite{franz_low_2009}. The sample is mainly bombarded with heavy ions such that unprotected carbon atoms are sputtered away. On the contrary, a high pressure gas of oxygen radicals without acceleration, as for example in a plasma asher (PA), will lead to a more chemical and undirected etching process relying on the oxidation of graphene \cite{franz_low_2009}.
However, in the rich literature of experiments with graphene nanoribbons etched by plasma, the technique used to pattern the devices is not always described in details. For instance, many papers mention "oxygen plasma etching" without clarifying if the plasma is produced in a RIE or a PA chamber \cite{han_energy_2007, chen_graphene_2007, todd_quantum_2008, murali_breakdown_2009, lian_quantum_2010, ryu_raman_2011, ribeiro_unveiling_2011, nakaharai_gate-controlled_2012, hollander_short-channel_2014}. These technical differences can induce structurally non-equivalent devices and an example of such a case is provided here. 
Using characterization methods like atomic force microscopy (AFM), Raman spectroscopy and electronic transport measurements, this work demonstrates that oxygen plasma ashing results in graphene ribbons with different properties than those obtained with argon and oxygen plasma RIE etching. 

\section{Materials and Methods}

\begin{figure}[t]
\centering
\includegraphics[width=8.5cm]{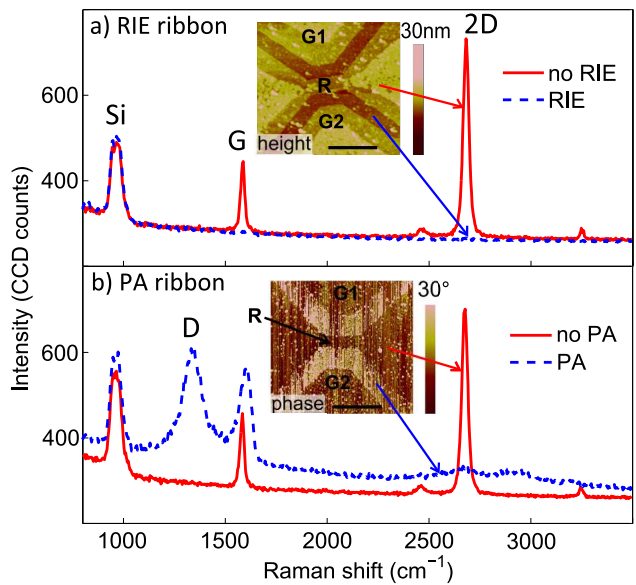}
\caption{Raman spectra taken in a region exposed to etching (dashed blue line) and in a protected region (continued red line) for the RIE ribbon (a) and the PA ribbon (b). AFM scans of the devices are displayed in the insets, with the letters 'R', 'G1' and 'G2' indicating the graphene ribbon, side-gate G$_{\rm 1}$ and side-gate G$_{\rm 2}$ respectively. The black scale bar is $500\,$nm. }
\label{fig1}
\end{figure}

Graphene flakes have been mechanically exfoliated on a doped Si/SiO$_{\rm 2}$ substrate. In two steps of electron beam lithography with a PMMA resist, the pristine flakes were first contacted with Cr/Au electrodes and then etched into the shape of a graphene ribbon with side-gates according to the structure presented in the insets of Fig.~\ref{fig1}. Two different etching processes were used: the first process is an O$_{\rm 2}$ plasma ashing step, performed in a Technics Plasma TePla100 asher at a pressure of $0.75\,$Torr and a power of $200\,$W for $100\,$s. The process time was chosen such that the ashed patterns in monolayer graphene were electrically insulating but their broadening (around $100\,$nm) remained acceptable for the design of nanostructures. The second process is an  Ar+O$_{\rm 2}$ reactive ion etching, performed in an Oxford Instruments PlasmaLab80Plus system at a pressure of $40\,$mTorr and a power of $35\,$W for ten seconds.  The time is chosen such that one layer of graphene is reliably etched. 

AFM images of the two ribbons, whose measurements are presented in this paper, are shown in the insets of Fig.~\ref{fig1}. The reactive ion etched ribbon (RIE ribbon) is the brighter central region in Fig.~\ref{fig1}a indicated with an R. It is about $155\,$nm wide and $360\,$nm long. The etched trenches, darker in the image, are about $130\,$nm wide which corresponds exactly to their designed width. The side-gates are indicated with the letters G1 and G2. They were patterned such that they are at a constant distance from the central ribbon and the graphene leads. The steps of different heights correspond to the SiO$_{2}$ substrate, the graphene flake and some residual EBL resist layer, which could not be completely removed using solvents. The plasma ashed ribbon (PA ribbon) is patterned into the same geometry. The ribbon can be seen as the darker central region in the inset of Fig.~\ref{fig1}b, indicated with an R. It is about $110\,$nm wide and $370\,$nm long, while the gaps between ribbon and side-gates (brighter regions) are about $200\,$nm wide, which is is about $100\,$nm wider than the designed pattern. An AFM phase image is depicted in Fig.~\ref{fig1}b, as this softer etching process does not create deep trenches. However, the different phase indicates that something else than pristine graphene resides in the unprotected regions. 

\section{Raman characterization}

In order to determine what materials are present after etching, Raman spectra of different regions on the two devices were recorded, as shown in Fig.~\ref{fig1}. For both devices, regions that were not exposed to any etching show the G and 2D peaks that are characteristic of graphene. The absence of a D peak confirms that the flake is mostly defect-free  \cite{ferrari_raman_2006}.  The full width at half maximum of the 2D peak of about $31\,\rm{cm}^{-1}$ indicates that the graphene flakes are monolayers. \\
The Raman spectra of the etched regions are however very different in both ribbons. As expected, only the Si peak remains in the reactive ion etched regions, meaning that the graphene has been entirely removed. Contrarily, the plasma ashed regions show a broadened G peak, a large D peak and an background signal is enhanced. This can be understood based on the two regimes of defective graphene found by Lucchese et al. \cite{lucchese_quantifying_2010}: they show that when defects are induced in a graphene flake using Ar$^+$ bombardment, a D-peak appears with increasing dose (first regime). Its intensity reaches a maximum and decreases again for larger doses (second regime). Other works \cite{childres_effect_2011,eckmann_probing_2012} also show that this D-peak evolution is accompanied by a decrease of the 2D peak intensity and a broadening of all peaks for both O$_{\rm 2}$ plasma and Ar exposures. In our case, the 2D peak has almost disappeared, the G peak is broadened and there is a large D peak. This reveals that the PA regions are deep in the second regime, called "highly disordered graphene": the structurally disordered parts of the graphene lattice dominate over the parts where the lattice is preserved. In other words, it is likely that there are small islands of intact graphene, but most of the material is so defective that it is not graphene anymore \cite{lucchese_quantifying_2010}. 
This "highly disordered regime" also exhibits electrically insulating behavior: we applied up to 6V to the side-gates of the PA ribbon and could not measure any leakage current to the ribbon. Therefore, it is possible that other experimental works involving nanostructures ashed with O$_{\rm 2}$ plasma have the same type of structure in the etched regions \cite{ryu_raman_2011, nakaharai_gate-controlled_2012}. \\
Eckmann et al. \cite{eckmann_probing_2012} further find that the different types of defects induced by high pressure O$_{\rm 2}$ plasma and by Ar bombardment result in different Raman signatures in the D' peak which is characteristic for intravalley scattering. Therefore, different types of defects will likely be present at the edges of the two devices. 
 
\section{Transport measurements}

We performed two-point conductance measurements through the ribbons at $T=1.6\,$K while tuning the voltage of the three different gates: the back-gate (BG) and the two graphene side-gates (G$_{\rm 1}$ and G$_{\rm 2}$). When sweeping the back-gate voltage at zero side-gate voltages, a region of suppressed conductance, called "transport gap" \cite{sols_coulomb_2007, molitor_transport_2009}, is observed for the PA ribbon. This is not the case for the RIE ribbon due to its larger width. However, by additionally tuning the side-gates, regions of suppressed conductance can be investigated in both ribbons.

\subsection{Transport in the RIE ribbon}
\begin{figure}[t]
\centering
\includegraphics[width=7.5cm]{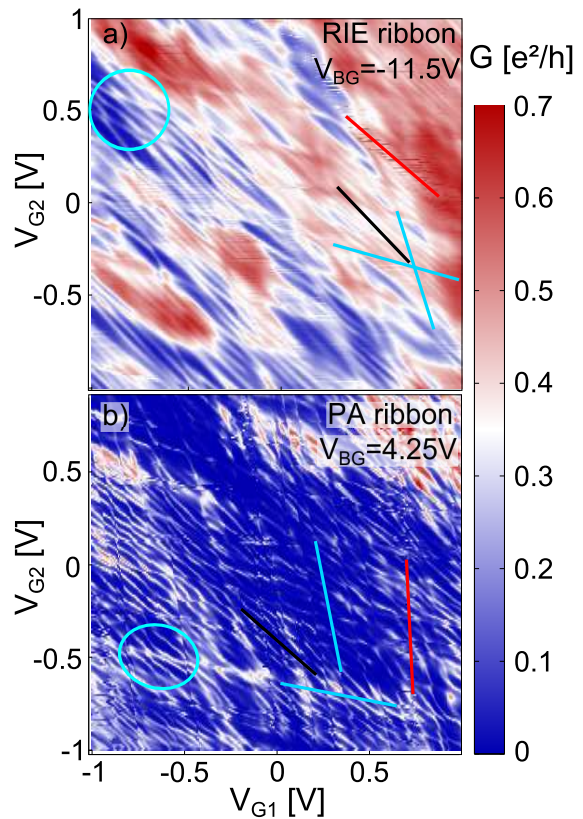}
\caption{ a) RIE ribbon and b) PA ribbon conductance as a function of both G$_{\rm 1}$ and G$_{\rm 2}$ voltages. An example of noisy line (red), steep resonance (black) and diagonal resonance (blue) is marked in each map. Light blue circles show examples of regions where anticrossings between resonances can be seen.}
\label{fig2}
\end{figure}

We first focus  on the transport in the RIE ribbon. 
Figure~\ref{fig2}a shows the conductance through the RIE ribbon as a function of the two side-gate voltages at a fixed back-gate voltage of $V_{\rm BG}= -8.8\,$V. Qualitatively, this measurement is very similar to other such maps on etched graphene nanoribbons \cite{todd_quantum_2008, bischoff_measuring_2015}. The conductance at a small bias of $500\,\mu$V varies from $0.005\,e^2/h$ up to $0.5\,e^2/h$ and features sharp resonances. Such resonances in graphene nanoribbons have been found to originate from Coulomb blockade occurring in sites of localized charge, similarly to a system of quantum dots \cite{sols_coulomb_2007,todd_quantum_2008,molitor_transport_2009}. In Fig.~\ref{fig2}a, most resonances have a slope of $-1$ (black line), indicating that both side-gates are equally well capacitively coupled to these sites. Since capacitance depends mostly on geometry, we can infer that the resonances roughly occur along the symmetry axis of the device between the two gates. Such resonances are generally not visible in bulk transport measurements and they can therefore be attributed to the ribbon at the center of the device. Conductance measurements as a function of bias and gate voltage give a charging energy of around $1\,$meV, as it is expected from the large ribbon width \cite{molitor_transport_2009, molitor_energy_2010}. \\
More generally, the capacitance ratio between the side-gates and localized states $\frac{C_{\rm G1-loc}}{C_{\rm G2-loc}}$ is related to the slope of resonances $\frac{\Delta V_{\rm G2}}{\Delta V_{\rm G1}}$ (see e.g. \cite{todd_quantum_2008, molitor_transport_2009}) according to 
\begin{equation}
\frac{C_{\rm G1-loc}}{C_{\rm G2-loc}}= -\frac{\Delta V_{\rm G2}}{\Delta V_{\rm G1}} \rm{.}
\end{equation}
In Fig.~\ref{fig2}a, additional resonances with slopes differing significantly from $-1$ can be observed (light blue lines), meaning that they are more influenced by one gate than by the other. Anticrossings between these resonances and the diagonal ones (see e.g. inside the light blue circles in Fig.~\ref{fig2}) imply that they are strongly coupled to the latter, at least capacitively \cite{van_der_wiel_electron_2002}. It is an indication that they also occur due to localized states, most probably located in the graphene. At the same time, their slopes show that their capacitance to one side-gate is significantly larger than the other (see Eq.~1). This suggests that they have to be off-centered (compared to the symmetry axis along the ribbon). Since defective, possibly functionalized edges are expected from the etching process, it is possible that these localized states sit at the edges of the graphene device. 
This finding is in agreement with previous work locating localized states in graphene nanoribbons at the edges of the ribbons or even at the edges of the leads \cite{bischoff_characterizing_2014}. \\
Further, a noisy line is found in Fig.~\ref{fig2}a (red line), also following a steep slope. By measuring the conductance as a function of time at gate voltages where such noise appears, a random telegraph signal is observed with a characteristic time of the order of $100\,$s. We therefore argue that a slow charge trap gates the ribbon conductance while the trap is loaded and unloaded, resulting in the observed noise.

\subsection{Comparison of transport in both ribbons}
Next, the transport in the PA and the RIE ribbons are compared. 
Figure~\ref{fig2}b shows a side-gate map of the conductance in the PA ribbon in a region of strongly reduced conductance. As expected, the transport in the narrower PA ribbon is more suppressed than in the wider RIE ribbon. Otherwise, the two maps are qualitatively similar: transport through the PA ribbon also features resonances of diagonal slope (black line) anticrossing with resonances tuned more strongly by one side gate compared to the other (light blue lines). Thus, as for the RIE ribbon, there are sites of localized charge coupled to each other, both at the center and towards the edges of the device.\\
To compare the two ribbons' transport properties, we now evaluate the slopes of the asymmetrically tuned resonances for side-gate maps where anticrossings can be seen. These slopes are found to vary widely as a function of the different side-gate and back-gate voltages. 
However, for all measured data and for all cool-downs, we find that the localized states close to gate G$_{\rm 1}$ give rise to slightly more extreme slopes in the case of the PA ribbon: they are ranging from $-3$ to $-7$ compared to a range of $-2$ to $-5$ for the RIE ribbon. The resonances more strongly tuned by gate G$_{\rm 2}$ give the same result (from $-1/2$ to $-1/6$ for the RIE ribbon and from $-1/2.5$ to $-1/8.5$ for the PA ribbon). \\
From a purely geometrical argument this is not expected: the PA ribbon is narrower and has wider trenches than the RIE ribbon. Thus, asymmetrically tuned resonances arising from the PA ribbon should have slopes closer to $-1$ compared to the ones in the RIE ribbon. The opposite finding can be explained if the localized states giving rise to these resonances are more off-centered in the PA device than they are in the RIE device. Indeed, these localized states could be ({\it i}) closer to the edges in the PA ribbon than in the RIE ribbon, ({\it ii}) extending further into the graphene leads adjacent to the PA ribbon than for the RIE ribbon, ({\it iii}) extending into the gap between ribbon and side-gates, where disordered graphene is present, or ({\it iv}) any combination thereof. \\
\begin{figure}[t]
\centering
\includegraphics[width=8.5cm]{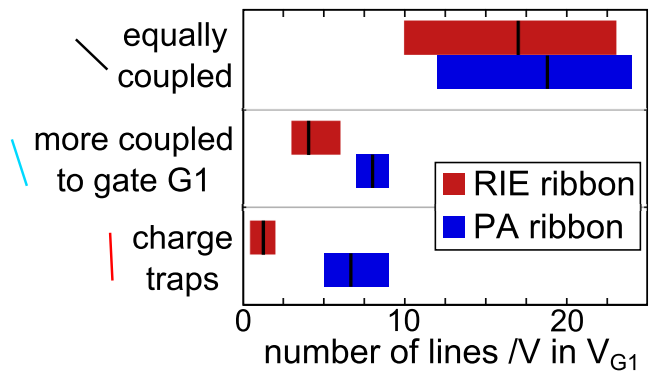}
\caption{Number of diagonal resonances, steep resonances and noisy lines per volt in side-gate maps in the region of suppressed conductance. The bars represent the spread between the minimal and maximal numbers found in the measurements, while the black mark is the average number.}
\label{fig3}
\end{figure}
Another difference between the two ribbons is the spacing of conductance resonances. In a quantum dot picture, the spacing between two Coulomb resonances in gate voltage $\Delta V_{\rm G}$ is directly related to the capacitance between the dot and the gate: $\Delta V_{\rm G} =e/C_{\rm dot-G}$. In a graphene ribbon, however, we expect the presence of several coupled sites of localized charge \cite{sols_coulomb_2007,todd_quantum_2008, molitor_transport_2009}. Thus, different resonances might be caused by different localized sites.\\
We therefore consider in Fig.~\ref{fig3} the inverse spacing $\rm{i.e.}$ the number of lines per volt for the resonances and the noisy lines. For each side-gate map measured in the region of suppressed conductance, the mean number of lines per volt is calculated. Since no significant effect of the temperature or cool-downs was found, this number is further averaged over at least $5$ side-gate maps from different cool-downs at $T=1.6\,$K and $T=4.2\,$K.
The color bars show the minimal and maximal values measured in these side-gate maps. For the resonances equally tuned by both side-gates and for both ribbons, the maximal number of resonances per volt is at least twice the minimum. This is the result of a changing line spacing in back-gate voltage. Thus, it is difficult to conclude on the relative size and number of localized states in the center of the two ribbons. More consistently% as a function of the back-gate voltage
, the PA ribbon transport suffers from $5$ times more instabilities and noisy lines than the RIE ribbon. This clear trend was also observed in another narrower PA ribbon on SiO$_2$ and in a wide PA ribbon on GaAs. This manifests a higher number of slow charge traps between the PA ribbon and its side-gates. \\
Finally, there is a tendency towards more non-diagonal resonances in the PA ribbon transport. Indeed, on average twice more of such resonances are observed than in the RIE ribbon transport. 
This is again not expected from a geometrical point of view: the PA trenches are wider than the RIE ones, so the side-gate capacitance to the ribbon and its edges should be smaller.  Hence, if the two ribbons had the same distribution of localized charges, one would expect a smaller number of resonances per volt for the PA ribbon. This opposite result indicates that the off-centered localized states in the PA device could be  ({\it i}) more numerous,  ({\it ii}) larger,  ({\it iii}) closer to the edges of the device,  ({\it iv}) extending in the gap between ribbon and side-gates, or ({\it v})  any combination thereof. \\
Including the Raman spectroscopy findings, we speculate that the off-centered localized states might include disordered graphene islands in the ashed patterns that are still connected to the device. These islands would  increase the effective width of the ribbon and thus enhance the side-gates influence. This would then explain why the PA ribbon features both more numerous and more asymmetrically tuned resonances than the RIE ribbon. Furthermore, the charge traps responsible for the instability and noise in the PA ribbon transport might arise from disordered graphene islands disconnected from the device.  This example and the noisy transport observed in the two other ribbons we fabricated this way show that in principle plasma ashing is less reliable to efficiently etch nanostructures in graphene.

\section{Conclusion}

Oxygen plasma ashing has the advantage of minimizing the implantation of defects in the substrate. However, in order to obtain nanodevices, short etching times are needed to prevent a broadening of the exposed patterns. This results in electrically insulating regions with residual carbon patches. The studied devices demonstrate that plasma ashing with low dose induces more instabilities and localized states in the graphene nanostructures than complete RIE etching. The former technique is therefore ill-suited in order to fabricate devices intended for transport measurements. A hard mask lithography process could allow applying higher doses while still using the less invasive ashing technique. This study also highlights the importance of choosing a suitable processing technique to achieve graphene devices with good electronic properties. 

\begin{acknowledgments}
The authors would like to thank Axel Eckmann, Florian Libisch and Aleksey Kozikov for helpful discussions. Financial support by the Marie Curie Initial Training Action (ITN) Q-NET 264034 and the National Center of Competence in Research on "Quantum Science and Technology" (NCCR QSIT) funded by the Swiss National Science Foundation is gratefully acknowledged.
\end{acknowledgments}

% Create the reference section using BibTeX:

\bibliography{Nanoribbons,Raman,addbib}

\end{document}